\documentclass[12pt]{article}
\usepackage{graphicx}
\usepackage{amsmath}
\begin{document}
\vspace*{4cm}
\begin{center}

\textbf{{\LARGE On the Casimir effect in the
microelectromechanical systems MEMS }}

\end{center}

\begin{center}

Janina Marciak-Kozlowska, Miroslaw Kozlowski$^{*}$\\
Institute of Electron Technology, Al. Lotnik\'{o}w 32/46, Warsaw
Poland

\end{center}

\vspace{1.5cm}
\hbox to 5cm{\hsize=5cm\vbox{\ \hrule}}\par
$^{*}$ Corresponding author, e-mail: MiroslawKozlowski@aster.pl

\vspace{2cm}
\begin{abstract}
In this paper the thermal transport phenomena in MEMS are
investigated. The thermal Klein-Gordon transport equation for
nanoscale structures is formulated and solved.


\end{abstract}

\newpage
\section{Introduction}
The Casimir effect is one of the most strinking consequences of
quantum electrodynamics~\cite{1}. The dependence of the ground
state energy of the electromagnetic field upon boundary conditions
gives rise to an observable force between macroscopic bodies. A
significant enhancement in the accuracy of measuring the Casimir
force has been achieved recently with experiments employing the
torsion pendulum and atomic force microscope (AFM)~\cite{2}.
Casimir effect investigations may open the way for experimental
observation of new fundamental forces arising from the
hypothetical extra dimension predicted by modern
theories~\cite{3}. However, to enable such studies it is crucial
to improve experimental techniques.

The Casimir force in addition to its fundamental interest also
plays an important role in the fabrication an operation of
microelectromechanical systems~(MEMS).

In this paper we describe the heat signaling in the simple
nanostructure the parrarel plates heated by attosecond laser
pulses. It will be shown that temperature field between plates
depends on the distance of the plates. As the result the
attosecond laser pulse can be used as the tool for the
investigation of Casimir effect on the performance of the MEMS.
\section{Repulsive quantum vacuum forces}
Vacuum energy is a consequence of the quantum nature of the
electromagnetic field, which is composed of photons. A photon of
frequency $\omega $
 has energy
$\hbar \omega $, where $\hbar $
 is Planck constant. The quantum vacuum can be interpreted as
the lowest energy state (or ground state) of the electromagnetic
(EM) field that occurs when all charges and currents have been
removed and the temperature has been reduced to absolute zero. In
this state no ordinary photons are present. Nevertheless, because
the electromagnetic field is a quantum system the energy of the
ground state of the EM is not zero. Although the average value of
the electric field $\left\langle E\right\rangle $
 vanishes in ground state, the Root Mean Square of the field
$\left\langle E^{2} \right\rangle $
 is not zero. Similarly the
$\left\langle B^{2} \right\rangle $
 is not zero. Therefore the electromagnetic field energy,
$\left\langle E^{2} \right\rangle +\left\langle B^{2}
\right\rangle $
 is not equal zero. A detailed theoretical calculation tells
that EM energy in each mode of oscillation with frequency $\omega$
 is $0.5\,\hbar \omega $, which equals one half of the amount of
energy that would be present if a single ``real'' photon of that
mode were present. Adding up $0.5\,\hbar \omega $
 for all possible modes of the electromagnetic field gives a
very large number for the vacuum energy $E_{0} $
 in the quantum vacuum
    \begin{equation}
    E_{0} =\sum\limits_{i}\frac{1}{2} \hbar \omega
    _{i}.\label{eq1}
    \end{equation}
     The
resulting vacuum energy $E_{0} $
 is \textit{infinity} unless a high frequency cut off is applied.\\
Inserting surfaces into the vacuum causes the modes of the EM to
change. This change in the modes that are present occurs since the
EM must meet the appropriate boundary conditions at each surface.
Surface alter the modes of oscillation and therefore the surfaces
alter the energy density corresponding to the lowest state of the
EM field. In actual practice the change in $E_{0} $
 is defined as follows
    \begin{equation}
    \Delta E_{0} =E_{0} -E_{S} ,\label{eq2}
    \end{equation}
 where $E_{0}$ is the energy in empty space and $E_{S} $
 is the energy in space with the surfaces, i.e.
    \begin{equation}
    \Delta E_{0} =\frac{1}{2} \sum\limits_{n}^{\rm \stackrel{empty}{\scriptscriptstyle space}}
    \hbar \omega _{n}  -\frac{1}{2} \sum\limits_{i}^{\rm\stackrel{surface}{\scriptscriptstyle present}}
    \hbar \omega _{i}.\label{eq3}
    \end{equation}
As an example let us consider a hollow conducting rectangular
cavity with sides \textit{a}$_{1,}$ \textit{a}$_{2}$,
\textit{a}$_{3}$. In that case for uncharged parallel plates with
area \textit{A} the attractive force between the plates
is,~\cite{4}
    \begin{equation}
    F_{att} =-\frac{\pi ^{2} \hbar c}{240d^{4} } A,\label{eq4}
    \end{equation}
     where \textit{d} is the distance between plates.
The force $F_{\rm att} $
 is called the parallel plate Casimir force, which was measured
in three different  experiments~\cite{5,7}.

 Recent calculations
show that for conductive rectangular cavities the vacuum forces on
a given face can be repulsive (positive), attractive (negative) or
zero depending on the ratio of the sides~\cite{8}.

 In
paper~\cite{5} the first measurement of repulsive Casimir force
was performed. For the distance (separation) $d\sim0.1\,$m
 the repulsive force is of the order of
$0.5\,\mu $N (micronewton) -- for cavity geometry. In March 2001,
scientist at Lucent Technology used attractive parallel plate
Casimir force to actuate a MEMS torsion device~\cite{7}. Other
MEMS (MicroElectroMechanical Systems) have been also
proposed~\cite{10}.
\section{Klein-Gordon thermal equation with Casimir force}
Standard Klein-Gordon equation reads:
    \begin{equation}
    \frac{1}{c^{2} } \frac{\partial ^{2} \Psi }{\partial t^{2} }
    -\frac{\partial ^{2} \Psi }{\partial x^{2} } +\frac{m^{2} c^{2}
    }{\hbar ^{2} } \Psi =0.\label{eq5}
    \end{equation}
In equation~(\ref{eq5}) $\Psi $
 is the relativistic wave function for particle with mass \textit{m}, \textit{c}
is the light velocity and $\hbar $
 is Planck constant. For massless particles
$m=0$  and Eq.~(\ref{eq5}) is the Maxwell equation for photons. As was
shown by Pauli and Weisskopf since relativistic quantum mechanical
equation had to allow for creation and anihilation of particles,
the Klein-Gordon describes spin -- 0 bosons.\\
In the monograph by two of us (Janina Marciak-Kozlowska and
Miroslaw Kozlowski)~\cite{11} the generalized Klein-Gordon thermal
equation was developed
    \begin{equation}
    \frac{1}{v^{2} } \frac{\partial ^{2} T}{\partial t^{2} }
     -\nabla^{2} T+\frac{m}{\hbar } \frac{\partial T}{\partial t}
    +\frac{2Vm}{\hbar ^{2} } =0\label{eq6}
    \end{equation}
In Eq.~(\ref{eq6}) \textit{T} denotes temperature of the medium
and \textit{v} is the velocity of the temperature signal in the
medium. When we extract the highly oscillating part of the
temperature field,
    \begin{equation}
    T=e^{-\,\frac{t\omega }{2} } u(x,t),\label{eq7}
    \end{equation}
where $\omega =\tau ^{-1} $ , and $\tau $
 is the relaxation time, we obtain from Eq. (3) (1D case)
    \begin{equation}
    \frac{1}{v^{2} } \frac{\partial ^{2} u}{\partial t^{2} }
    -\frac{\partial ^{2} u}{\partial x^{2} }
    +qu(x,t)=0,\label{eq8}
    \end{equation}
where
    \begin{equation}
    q=\frac{2Vm}{\hbar ^{2} } -\left( \frac{mv}{2\hbar } \right) ^{2}
    \label{eq9}
    \end{equation}
When $q>0$
 equation~(\ref{eq8}) is of the form of the Klein-Gordon equation
in the potential field $V(x,t)$. For $q<0$
 Eq.~(\ref{eq8}) is the modified Klein-Gordon equation. The discussion
of the physical properties of the solution of equation~(\ref{eq8})
can be find in~\cite{8}.

In the paper we will study the heat signaling in the medium
excited by ultra-short laser pulses, $\Delta t<\tau $. In that
case the solution of Eq.{\nobreakspace}(1) can be approximated as
    \begin{equation}
    T(x,t)\cong u(x,t)\quad \quad for \quad \quad \Delta
    t\ll\tau.\label{eq10}
    \end{equation}
Considering the existence of the attosecond laser with $\Delta
t\approx 1\,$as$=10^{-18} $s, Eq.~(\ref{eq8}) describes the heat
signaling for thermal energy transport induced by ultra-short
laser pulses. In the subsequent we will consider the heat
transport when \textit{V} is the Casimir potential. As was shown
in paper~\cite{7} the Casimir force, formula~(\ref{eq4}), can be
repulsive sign $(V)=+1$
 and attractive sign $(V)=-1$. For attractive Casimir force,
$V<0,\;q<0$  (formula (5)) and equation~(\ref{eq4})  is the
modified K-G equation. For repulsive Casimir force $V>0$
 and \textit{q} can be positive or negative.

As was shown by J.~Maclay~\cite{8} for different shapes of
cavities the vacuum Casimir force can be changes the sign. In the
subsequent we consider the propagation of thermal wave in the
geometry described in Fig.~1. In Fig.~2(a) the value of the
parameter \textit{q} as the function of \textit{d} -- distance
between the plates and $\frac{v}{c} $
 the ratio of the \textit{v --} thermal wave velocity and \textit{c --} the
light velocity is presented. Considering that $v\approx 10^{-2}
c$, from Fig. 2(b) we conclude that \textit{q} change the sign for
\textit{d}{\nobreakspace}={\nobreakspace}0.759{\nobreakspace}nm.
For Cauchy initial condition:
    $$u(x,0)=0,\quad \quad \frac{\partial u(x,0)}{\partial t}
    =f(x)$$
the solution of Eq.{\nobreakspace}(8) has the form
    \begin{equation}
    u(x,t)=\frac{1}{2v} \int\limits_{x-vt}^{x+vt}f(x)J_{0}
    \sqrt{q\left( v^{2} t^{2} -\left( x-\zeta \right) ^{2} \right) }
    d\zeta \quad \quad for\quad \quad q>0 \label{eq11}
    \end{equation}
and
    \begin{equation}
    u(x,t)=\frac{1}{2v} \int\limits_{x-vt}^{x+vt}f(x)I_{0}
    \sqrt{-q\left( v^{2} t^{2} -\left( x-\zeta \right) ^{2} \right) }
    d\zeta \;\quad \quad for\quad \quad q<0.\label{eq12}
    \end{equation}
In the subsequent we choose as the initial distribution
temperature, \textit{f}(\textit{x}), viz.
    $$f(x)=\exp\left[ -x^{2}
\right] .$$ In Fig.{\nobreakspace}3(a, b) we present the solution
of equation{\nobreakspace}(8) for $q=0$. In
Fig.{\nobreakspace}3(a) $u(x,t)=T(x,t)$
 is calculated as the solution of the modified K-G. Both solution
are the same shape due to the fact that $q=0$. In
Fig.{\nobreakspace}4(a, b) the solution of K-GE and MK-G equation
i.e. for $q>0$
 and
$q<0$
 are presented for different value of \textit{q}.

\newpage

\begin{figure}
\centering\includegraphics[height=12cm]{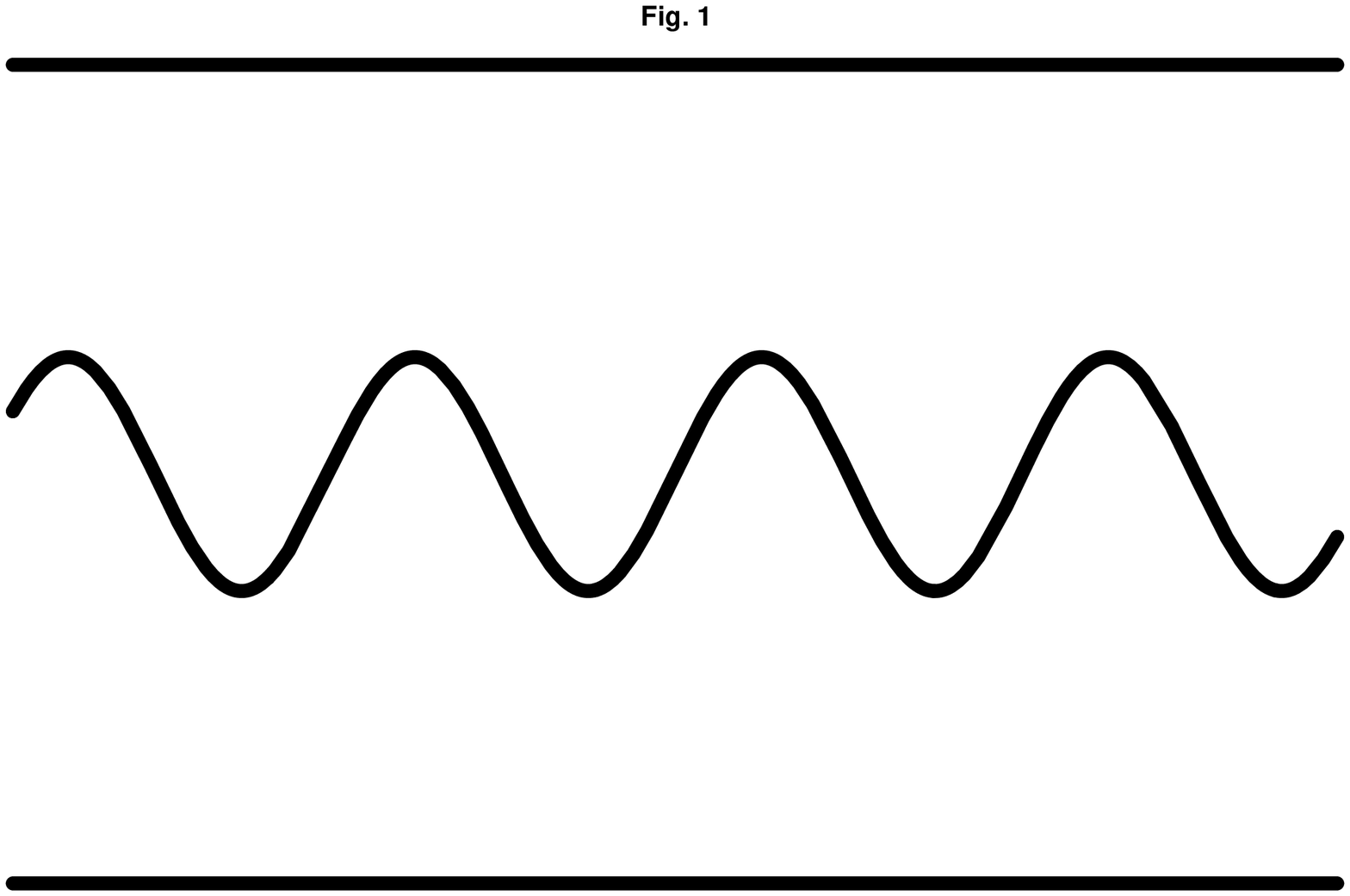} \caption{The
thermal wave and parallel plates}
\end{figure}
\begin{figure}
\centering\includegraphics[width=10cm]{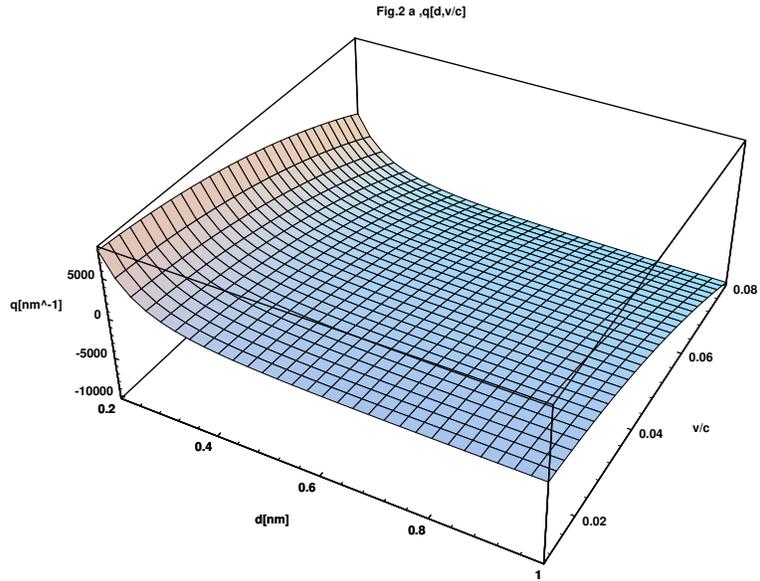}
\centering\includegraphics[width=10cm]{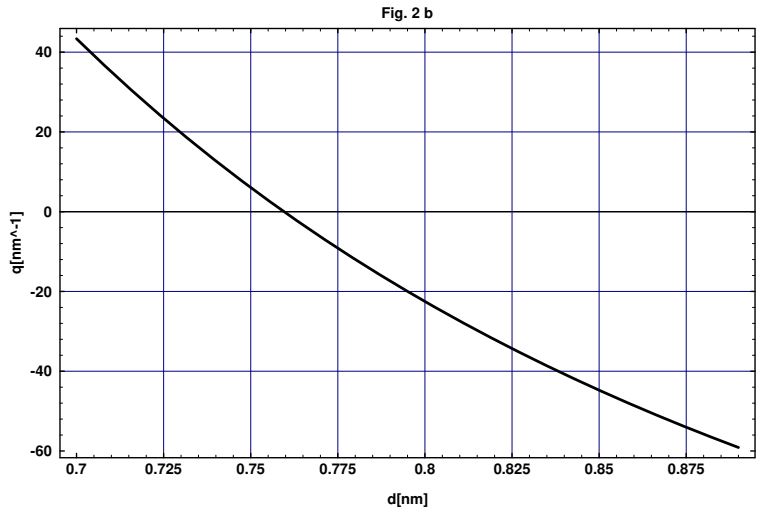} \caption{The
\textit{q} parameter as the function of \textit{d} and
\textit{v}/\textit{c}, (b) The \textit{q} parameter as the
function of \textit{d}, for $v=10^{-2} c$}
\end{figure}
\begin{figure}
\centering\includegraphics[width=10cm]{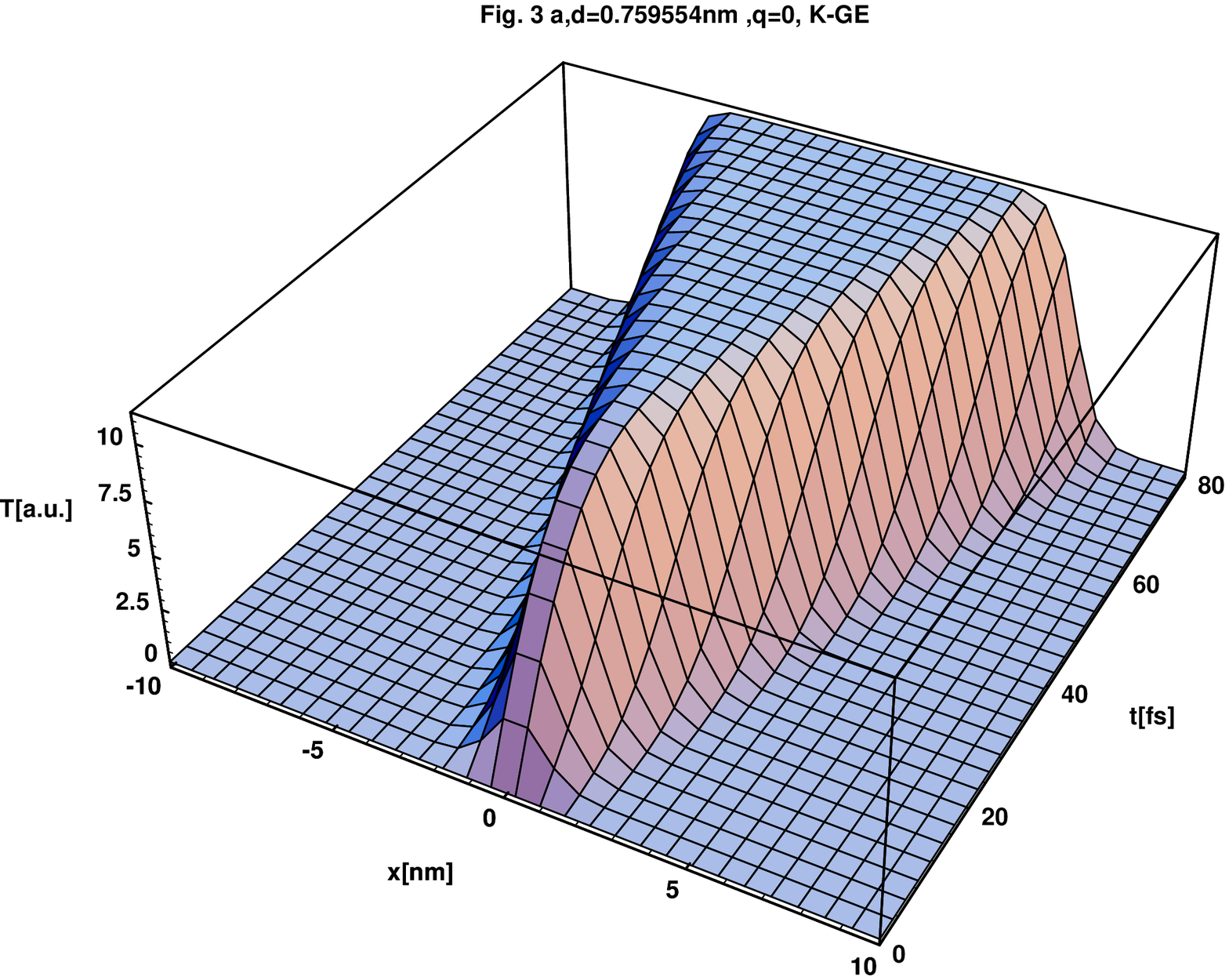}
\centering\includegraphics[width=10cm]{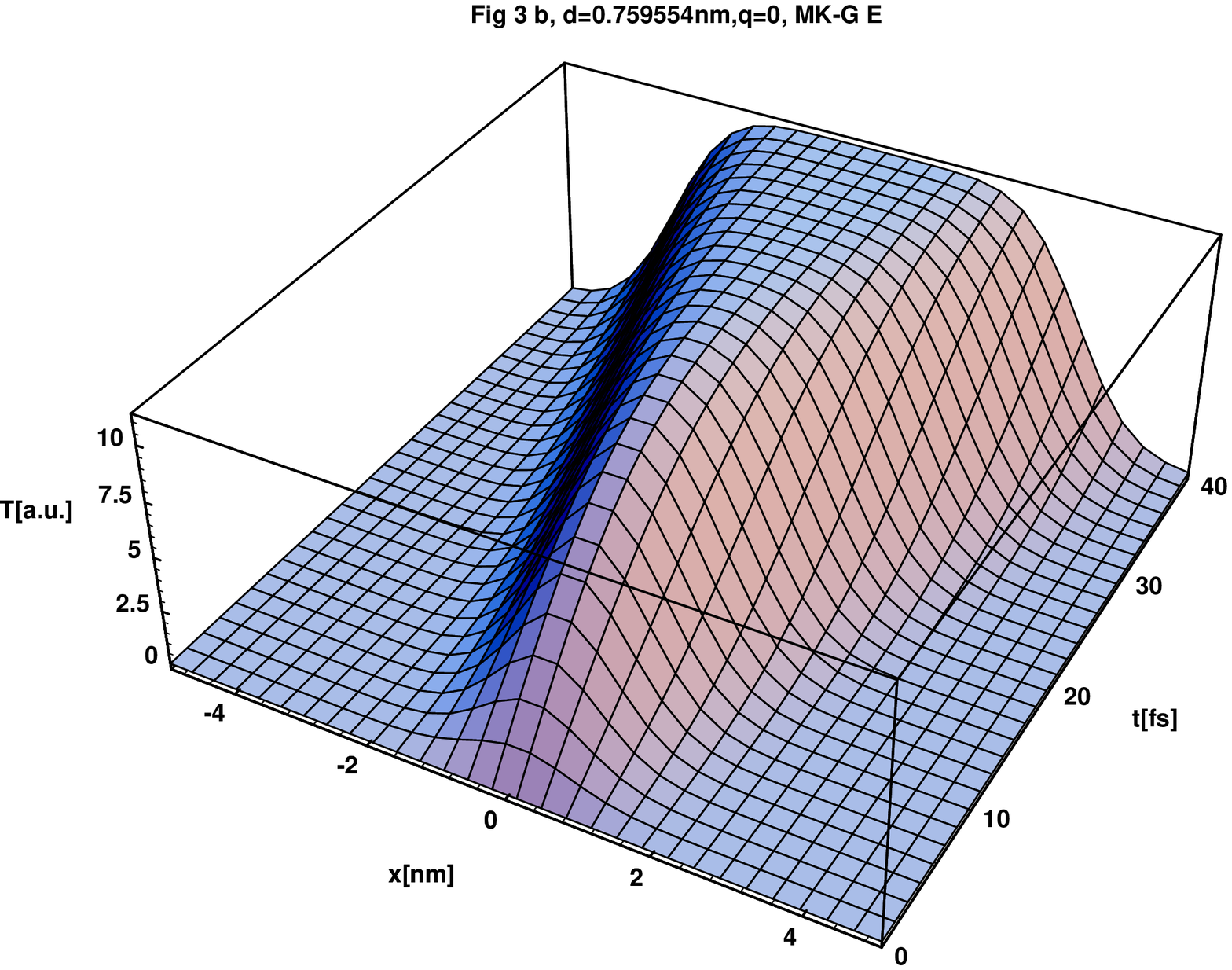} \caption{The
solution of equation~(\ref{eq8}) for \textit{d}~=~0.759~nm,
\textit{q}~=~0. (b) The solution of equation MK-GE (8) for
\textit{d}~=~0.759~nm, \textit{q}~=~0. }\end{figure}
\begin{figure}
\centering\includegraphics[width=10cm]{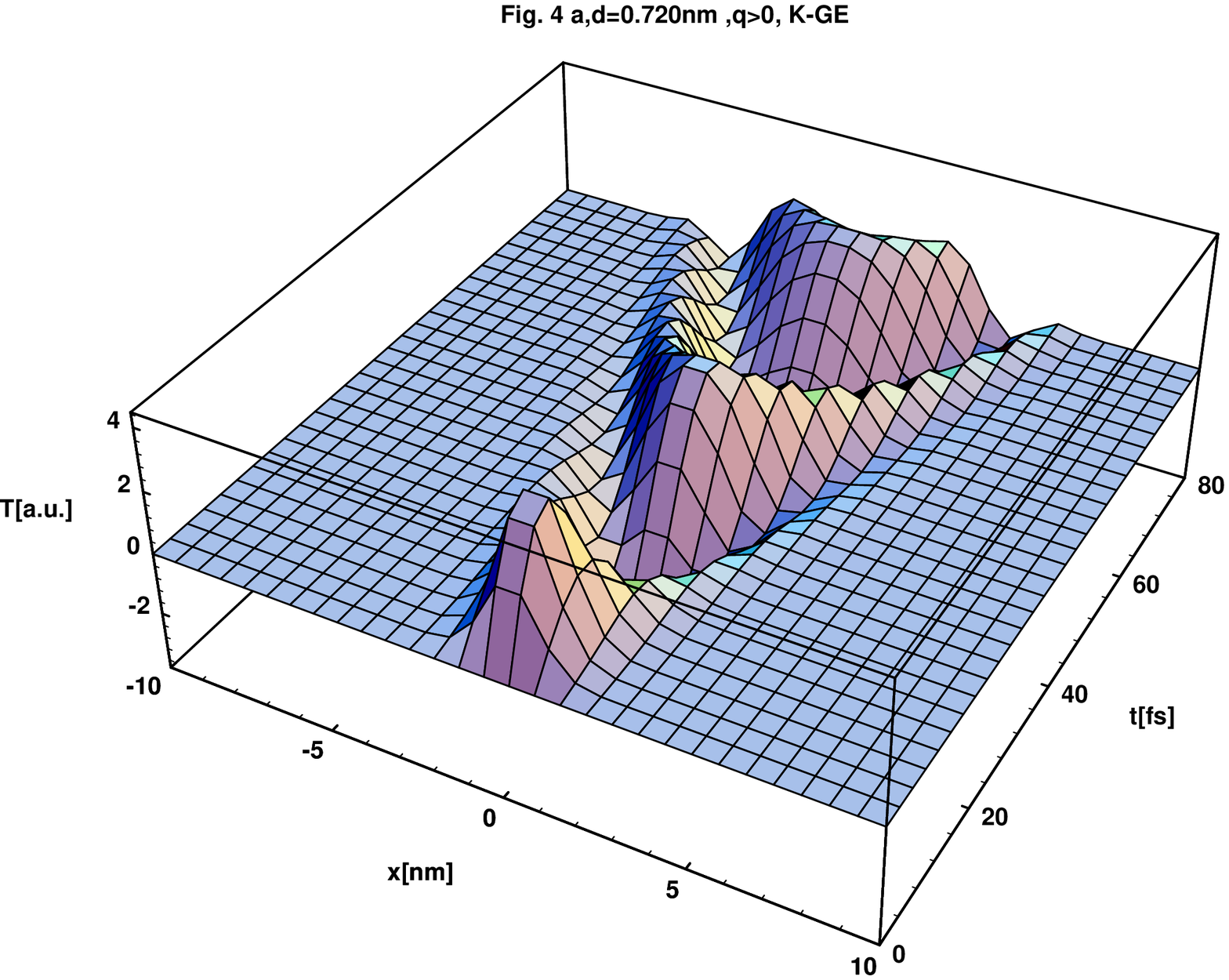}
\centering\includegraphics[width=10cm]{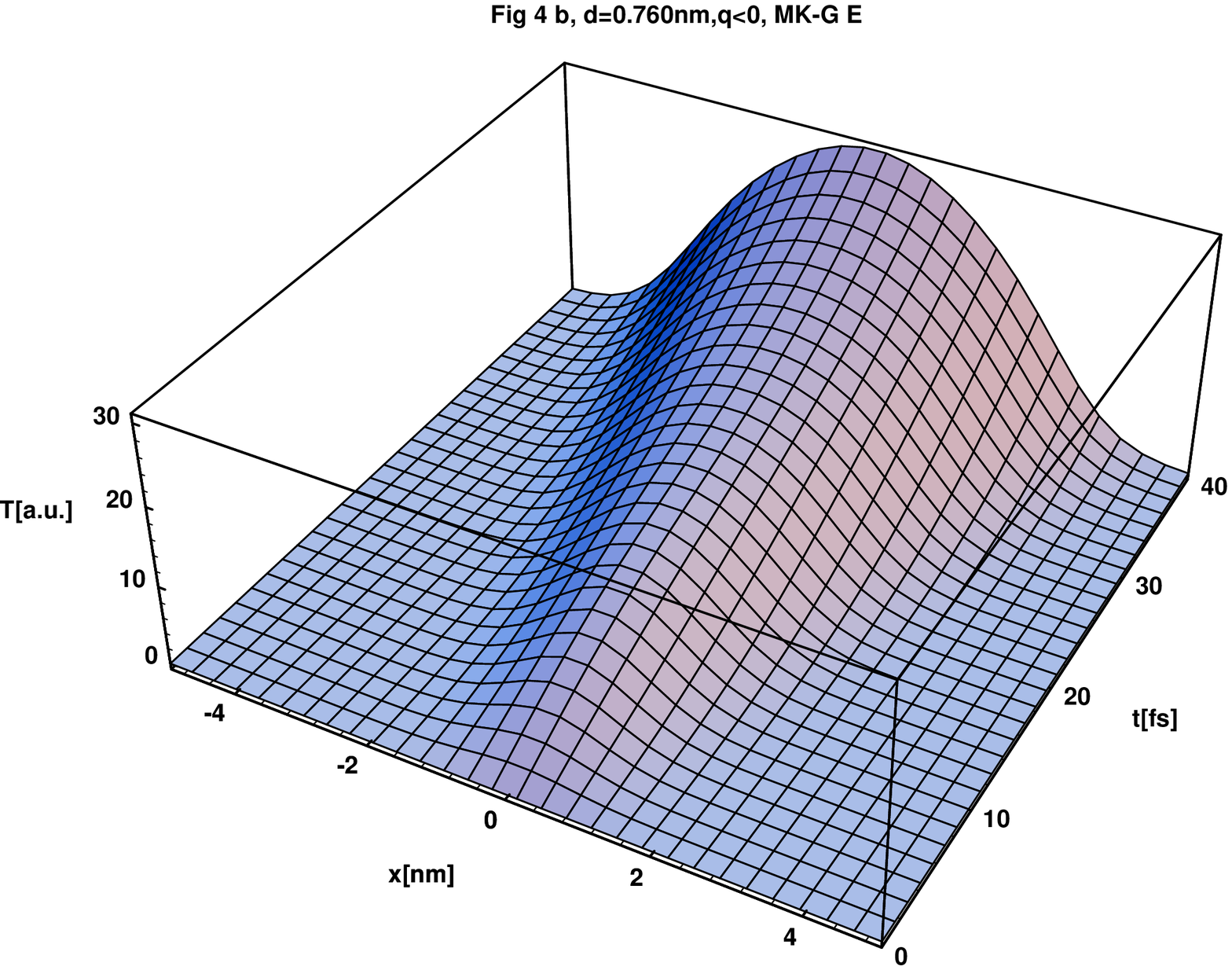} \caption{(a)
The solution of equation K-GE (8) for
\textit{d}{\nobreakspace}={\nobreakspace}0.720,
\textit{q}{\nobreakspace}\texttt{>}{\nobreakspace}0. (b) The
solution of equation MK-GE (8) for
\textit{d}{\nobreakspace}={\nobreakspace}0.760,
\textit{q}{\nobreakspace}\texttt{<}{\nobreakspace}0}
\end{figure}

\newpage
\begin{thebibliography}{99}
\bibitem{1}K. A. Milton, \textit{The Casimir effect}, World Scientific,
2001.
\bibitem{2}U. Mohideen, A. Roy, \textit{Phys. Rev. Lett.},
\textbf{81}, (1998), p. 4549.
\bibitem{3} J. C. Long et al., \textit{Nucl. Phys.},
\textbf{B539}, (1999), p.~23.
\bibitem{4}L. Brown and J. Maclay, \textit{Phys. Rev}.,
\textbf{184,} (1969), p.~1272.
\bibitem{5}S. Lamoreaux, \textit{Phys. Rev. Lett}., \textbf{78}, (1997), p.~5.
\bibitem{7}H. B. Chan et al., \textit{Science}, \textbf{291}, (2001), p. 1941.
\bibitem{8}J. Maclay, \textit{Phys. Rev}., \textbf{A61}, (2000), p.
052110.
\bibitem{9}J. Maclay et al., AIAA/ASME/SAE/ASEE, AIAA-2001-3359.
\bibitem{10}M. Serry et al., J. \textit{Microelectromechanical}
\textit{System}, \textbf{4,}
(1995), p. 193,
\bibitem{11}M. Kozlowski, J. Marciak-Kozlowska, \textit{From quarks o
bulk matter}, Hadronic Press 2001, USA.
\end{thebibliography}
\end{document}